\catcode`\@=11 % This allows us to modify PLAIN macros.
\def\lsim{\mathrel{\mathpalette\@versim<}}
\def\gsim{\mathrel{\mathpalette\@versim>}}
\def\@versim#1#2{\vcenter{\offinterlineskip
        \ialign{$\m@th#1\hfil##\hfil$\crcr#2\crcr\sim\crcr } }}
\catcode`\@=12 % at signs are no longer letters
\def\Tr{{\rm Tr\,}}
\def\D{{\partial}}

\documentstyle[ichep,12pt]{article}
\title{\normalsize OBSERVING ELECTROWEAK SYMMETRY BREAKING \\
       AT THE SSC\thanks{This work was supported
       in part by NSF grant PHY-90-96198.  It was
       done in collaboration with V.~Barger, K.~Cheung,
       J.~Gunion, T.~Han, G.~Ladinsky, R.~Rosenfeld and
       C.-P.~Yuan.}}

\author{Jonathan A. Bagger\\
        Department of Physics and Astronomy\\
        The Johns Hopkins University\\
        Baltimore, MD  21218}

\begin{document}
\finalcopy

\maketitle

\def\abstract#1{\vbox{
\twocolumn[\vbox
{\hsize\textwidth\parindent0pt\leftskip.75in\rightskip.75in
\vskip\baselineskip
#1\par\vfil\vskip\baselineskip}
]
}}

\abstract{In this talk we survey the SSC signals and backgrounds
for the physics of electroweak symmetry breaking.  We study the
process $pp \rightarrow WWX$ and compute the
rate for the ``gold-plated''
signals $W^\pm \rightarrow \ell^\pm \nu$ and $Z \rightarrow
\ell^+\ell^-$
$(\ell = e,\mu)$ for a wide variety of models.   We use a forward jet
tag and central jet veto to suppress the standard-model backgrounds.
In this way we estimate the SSC sensitivity to the physics of
electroweak symmetry breaking.}

\vskip-1pc
\onehead{INTRODUCTION}

During the past decade, particle physics passed from
triumph to triumph.  The discovery of the $W$ and the $Z$
demonstrated that the gauge structure of the standard
model is correct.  Precision measurements from LEP now
indicate that the top-quark mass is about 150 GeV, and
when top is discovered, the picture will be complete.

Or will it?  Consider a theory with the known particles:
quarks, leptons,
gluons, the photon and the $W$ and $Z$.  Then compute the
scattering amplitude for two longitudinally-polarized $W$
particles.  As shown by Lee, Quigg and
Thacker,$^1$ the amplitude diverges with energy, and
(perturbative) unitarity is violated below 2 TeV in
the center of mass.  Clearly, something must happen before
then, and the SSC must be ready to find it.$^2$

On general grounds, we know that whatever unitarizes
the $W_LW_L$ scattering amplitude
must also be responsible for giving mass
to the $W$ and $Z$.  Present experimental results\break
\newpage
\vglue2pt
\noindent
shed little light on the issue.
The fact that $M_W \simeq M_Z
\cos\theta$ suggests that electroweak symmetry
breaking respects a global symmetry $G \supseteq SU(2)_L
\times SU(2)_R$, spontaneously broken to $H \supseteq
SU(2)_V$.  We will use this unbroken $SU(2)$ ``isospin''
symmetry to organize our thinking about the physics of
electroweak symmetry breaking.

\onehead{THE STANDARD MODEL}

In the standard model, the $W_LW_L$ scattering amplitudes are
unitarized by exchange of a spin-zero resonance, the Higgs
particle $H$.  The Higgs is contained in a complex scalar
doublet, $\Phi\ =\ (v + H) \exp(2i w^a \tau^a/v)$, whose
four components split into a triplet $w^a$ and a singlet $H$
under isospin.  The $w^a$ are the Goldstone bosons
that give mass to the $W$ and $Z$, while the singlet is the Higgs
particle $H$.

The standard-model Higgs potential is invariant under an $SU(2)_L
\times SU(2)_R$ symmetry,
\begin{equation}
\Phi\ \rightarrow\ L\,\Phi\,R^\dagger\ ,
\end{equation}
\noindent
with $L,R \in SU(2)$.  The vacuum expectation value
$\langle\Phi\rangle = v$ breaks the symmetry to the
diagonal $SU(2)$.  In the perturbative limit, it also
gives mass to the Higgs.

\onehead{BEYOND THE STANDARD MODEL}

There are many other alternatives that might
describe the physics of electroweak symmetry breaking.
In this talk we will study a variety of different
models, each of which is completely consistent with
all the data to date (including that from the $Z$).  The
models give an idea of the range of physics that might
be seen at the SSC.

The first major distinction between the models is whether
or not they are resonant in the $W_LW_L$ channel.  If they are
resonant, the models are classified by the spin and isospin
of the resonance.  If they are not, the analysis is more
subtle, and we shall see that all possibilities can be
described in term of two parameters.
In what follows, we will restrict
our attention to nonresonant models, and to models
with spin-zero, isospin-zero resonances (like the Higgs),
and spin-one, isospin-one resonances (like the techni-rho).

\twohead{Spin-zero, Isospin-zero Resonances}

1)  $O(2N).$
The first model we discuss represents an attempt to
describe the standard-model Higgs in the nonperturbative
domain.  In the perturbatively-coupled standard model,
the mass of the Higgs is proportional to the square root
of the scalar self-coupling $\lambda$.  Heavy Higgs
particles correspond to large values of $\lambda$.
For $M_H \gsim$ 1 TeV, naive
perturbation theory breaks down.

One possibility for exploring the nonperturbative regime
is to exploit the isomorphism between $SU(2)_L \times SU(2)_R$
and $O(4)$.  Using a large-$N$ approximation, one can
solve the $O(2N)$ model for all values of $\lambda$,
to leading order in $1/N$.  The resulting scattering
amplitudes$^3$ can be parametrized by the scale $\Lambda$ of
the Landau pole.  Large values of $\Lambda$ correspond
to small couplings $\lambda$ and relatively light Higgs
particles. In contrast, small values of $\Lambda$ correspond
to large $\lambda$ and describe the nonperturbative regime.
In this talk we will take $\Lambda = 3$ TeV as a caricature
of the strongly-coupled standard model.

2) {\it Scalar.}
The second model describes the low-energy regime of a
technicolor-like model whose lowest resonance is a
techni-sigma.  The effective Lagrangian for such a resonance
can be constructed using the techniques of Callan, Coleman,
Wess and Zumino.$^4$  The resulting Lagrangian is consistent
with the chiral symmetry $SU(2)_L \times SU(2)_R$,
spontaneously broken to the diagonal $SU(2)$.

In this approach, the basic fields are $\Sigma = \exp
(2 i w^a \tau^a/v)$ and a scalar $S$.  These fields
transform as follows under $SU(2)_L \times SU(2)_R$,
\begin{eqnarray}
\Sigma & \rightarrow & L\,\Sigma\,R^\dagger\ , \nonumber \\
S & \rightarrow & S\ .
\end{eqnarray}
To the order of interest, the Lagrangian
contains just two parameters, which we can take to
be the mass and the width of the $S$.  In what follows,
we will choose $M_S = 1.0$ TeV, $\Gamma_S = 350$
GeV.  These values give unitary scattering amplitudes
up to 2 TeV.

\twohead{Spin-one, Isospin-one Resonances}

1)  {\it Vector.}
This model provides a relatively model-independent
description of the techni-rho resonance that arises in
most technicolor theories.$^5$  As above, one can use the
techniques of CCWZ to construct the effective Lagrangian.
The basic fields are $\xi = \exp(i w^a \tau^a/v)$ and
a vector $\rho_\mu$, which transform as follows under
$SU(2)_L \times SU(2)_R$,
\begin{eqnarray}
\xi & \rightarrow & L\,\xi\,U^\dagger\ =\ U\,\xi\,R^\dagger\ ,
\nonumber \\
\rho_\mu & \rightarrow & U\rho_\mu\,U^\dagger + i g^{\prime
\prime-1}\, U \partial_\mu U^\dagger\ ,
\end{eqnarray}
\noindent
where $U(L,R,\xi) \in SU(2)$.

\begin{table}[t]
\begin{center}\vspace{-.19in}
\begin{tabular}{ l l }
\multicolumn{2}{l}{Table~1. Cuts, tags and vetos, by mode.}\\[2mm]
\hline\hline
$W^+ W^-$ Basic cuts & Tag and Veto \\
\hline
 $| y_{\ell} | < 2.0 $  &
 $E_{tag} > 3.0\ {\rm TeV}$  \\
 $P_{T,\ell} > 100\ {\rm GeV}$  &
 $3.0 < \eta_{tag} < 5.0$  \\
 $\Delta P_{T,\ell\ell} > 200\ {\rm GeV}$  &
 $P_{T,tag} > 40\ {\rm GeV}$  \\
 $\cos\phi_{\ell\ell} < -0.8$  &
 $P_{T,veto} > 60\ {\rm GeV}$  \\
 $M_{\ell\ell} > 250\ {\rm GeV}$  &
 $ | \eta_{veto} | < 3.0$  \\[2mm]
\hline
$Z Z$ Basic cuts & Tag only\\
\hline
 $| y_{\ell} | < 2.5 $  &
 $E_{tag} > 1.0\ {\rm TeV}$   \\
 $P_{T,\ell} > 40\ {\rm GeV}$  &
 $3.0 < \eta_{tag} < 5.0$   \\
 $P_{T,Z} > {1\over4} \sqrt{M^2_{ZZ} - 4 M^2_Z}$  &
 $P_{T,tag} > 40\ {\rm GeV}$   \\
 $M_{ZZ} > 500\ {\rm GeV}$  & \\[2mm]
\hline
$W^+ Z$ Basic cuts & Tag and Veto\\
\hline
 $| y_{\ell} | < 2.5 $  &
 $E_{tag} > 2.0\ {\rm TeV}$  \\
 $P_{T,\ell} > 40\ {\rm GeV}$  &
 $3.0 < \eta_{tag} < 5.0$  \\
 $P_{T, miss} >  75\ {\rm GeV}$  &
 $P_{T,tag} > 40\ {\rm GeV}$   \\
 $P_{T,Z} > {1\over4} M_T{}^*$ &
 $P_{T,veto} > 60\ {\rm GeV}$  \\
 $M_T > 500\ {\rm GeV}$ &
 $ | \eta_{veto} | < 3.0$  \\[2mm]
\hline
$W^+ W^+$ Basic cuts & Veto only \\
\hline
 $| y_{\ell} | < 2.0 $   &
 $P_{T,veto} > 60\ {\rm GeV}$  \\
 $P_{T,\ell} > 100\ {\rm GeV}$   &
 $ | \eta_{veto} | < 3.0$  \\
 $\Delta P_{T,\ell\ell} > 200\ {\rm GeV}$   & \\
 $\cos\phi_{\ell\ell} < -0.8$  & \\
 $M_{\ell\ell} > 250\ {\rm GeV}$  & \\
\hline\hline
\multicolumn{2}{l}{{}$^*$  $M_T$ is the cluster transverse
mass.$^5$}\\
\end{tabular}
\end{center}
\end{table}

For the processes of interest, the effective Lagrangian
again depends on just two couplings, the mass and the
width of the resonance.  In what follows we will choose
$M_\rho = 2.0$ TeV, $\Gamma_\rho = 700$ GeV and
$M_\rho = 2.5$ TeV, $\Gamma_\rho = 1300$ GeV.
These values preserves unitarity up to 3 TeV.

\twohead{Nonresonant models}

The final models we consider are
nonresonant at SSC energies.  In this case
the new physics contributes to the effective Lagrangian
in the form of higher-dimensional operators built
from the Goldstone fields.  To order $p^2$ in the
energy expansion, only one operator contributes, and
its coefficient is universal.  To order $p^4$, there are
just two additional
operators that contribute to $W_LW_L$
scattering.$^6$  They are
\begin{eqnarray}
{\cal L}^{(4)} &= &
{L_1\over 16\pi^2}\,\bigg(\Tr \D_\mu \Sigma
\D_\mu \Sigma^\dagger\bigg)^2\ \nonumber \\
&& +\ {L_2\over 16\pi^2}\,\bigg(\Tr \D_\mu \Sigma
\D_\nu \Sigma^\dagger\bigg)^2\ .
\end{eqnarray}
\noindent
The coefficients $L_1$ and $L_2$ contain all information
about the new physics.

The difficulty with this approach is that at SSC energies,
the scattering amplitudes violate unitarity between 1 and 2
TeV.  This is an indication that new physics is near, but
there is no guarantee that new resonances lie within the
reach of the SSC. We choose to treat the uncertainties of
unitarization in three ways:

1)  {\it LET CG.}
We take $L_1 = L_2 = 0$, and cut off the partial wave
amplitudes when they saturate the unitarity bound.
This is the original model considered by Chanowitz and
Gaillard.$^7$

2)  {\it LET K.}
We take $L_1 = L_2 = 0$, and unitarize with a K-matrix.

3)  {\it Delay K.}
We take $L_1 = -0.26$ and $L_2 = 0.23$, a choice that
preserves unitarity up to 2 TeV.  Beyond that scale, we
unitarize the scattering amplitudes with a K-matrix.$^8$

These three models describe possible nonresonant new physics
at the SSC.

\onehead{SIGNALS AND BACKGROUNDS}

In the rest of this talk we will focus on SSC signals and
backgrounds for the process $pp \rightarrow WWX$.  We will
concentrate on the ``gold-plated'' decays
$W^\pm \rightarrow \ell^\pm \nu$ and $Z \rightarrow \ell^+
\ell^-$, for $\ell = e,\mu$, in each of the final states
$W^+W^-$, $W^+ Z$, $ZZ$ and $W^+W^+$.

We will take the signal to be the process $pp \rightarrow
W_LW_LX$ because the longitudinal $W$'s couple most
strongly to the new physics.  We will take $pp \rightarrow
W_LW_TX$ and $pp \rightarrow W_TW_TX$ to be the background.
These processes are dominated by diagrams that do not depend
on the new physics, so we will
represent the background by the standard model with a light
Higgs (of mass 100 GeV).
The difference between this and the true background is
negligible at the energies we consider.

We will simplify our calculations by using the electroweak
equivalence theorem,$^{1,7}$ which lets us replace the longitudinal
vector bosons by their corresponding would-be Goldstone
bosons.  We will also use the effective $W$ approximation$^9$
to connect the $W_LW_L$ subprocesses to the $pp$
initial state.

In the $W^+W^-$, $W^+ Z$ and $ZZ$ channels, the final
states of interest are dominated by glue-glue and
$q \bar q$ scattering.$^{10}$  We suppress these contributions
by requiring a tag on the forward jet$^{11}$ associated with
an initial-state $W$.
In the $W^+W^-$, $W^+ Z$ and $W^+W^+$ channels, there
is a residual background from top decay that we suppress
by requiring a central jet veto.$^{12}$  The combination
of a forward jet tag and central jet veto is very effective
in reducing the background in all charge channels.

The precise cuts we use are summarized in Table 1.
In all channels, the dominant residual background is transverse
electroweak, followed by $q \bar q$ annihilation and top decay.

Because we use the effective $W$ approximation for our signal,
we can only estimate the effects of the tag and veto.  Therefore
we have used the exact standard-model calculation with a 1 TeV
Higgs to derive efficiencies for the tag and veto.  These
efficiencies are then applied to the effective $W$ calculations
to estimate the rate for each signal.  The results for the
signals and backgrounds are collected in Table 2.

\begin{table*}[t]
\begin{center}\vspace{-.19in}
\begin{tabular}{ l | c  c  c  c  c  c  c  c  c }
\multicolumn{10}{l}{Table~2. Event rates per SSC-year, assuming $m_t
= 140$ GeV,
$\sqrt{s} = 40$ TeV, and an annual}\\
\multicolumn{10}{l}{\hspace{44pt} luminosity of $10^4$
pb$^{-1}$.}\\[2mm]
\hline\hline
$W^+ W^-$ & Bkgd. & SM & Scalar & $O(2N)$ & Vec 2.0 & Vec 2.5 & LET
CG &
   LET K & Delay K  \\
\hline
$M_{\ell\ell} > 0.25$ & 9.1 & 59 & 31 & 26 & 12 & 10 & 12 & 10 &
9.7  \\
$M_{\ell\ell} > 0.5$ & 5.0 & 31 & 20 & 16 & 9.3 & 7.4 & 9.3 & 7.0
& 6.9  \\
$M_{\ell\ell} > 1.0$ & 0.9 & 2.0 & 0.6 & 1.5 & 3.6 & 2.6 & 2.9 &
1.9 & 2.5  \\[2mm]
\hline
$W^+ Z$ & Bkgd. & SM & Scalar & $O(2N)$ & Vec 2.0 & Vec 2.5 & LET CG
&
   LET K & Delay K  \\
\hline
$M_T > 0.5$ & 2.5 & 1.3 & 1.8 & 1.5 & 9.6 & 6.2 & 5.4 & 4.7 & 5.5  \\
$M_T > 1.0$ & 0.9 & 0.6 & 0.9 & 0.8 & 8.2 & 4.8 & 4.0 & 3.3 & 4.3  \\
$M_T > 1.5$ & 0.3 & 0.2 & 0.3 & 0.3 & 5.9 & 3.4 & 2.5 & 1.8 & 2.9
\\[2mm]
\hline
$Z Z$ & Bkgd. & SM & Scalar & $O(2N)$ & Vec 2.0 & Vec 2.5 & LET CG &
   LET K & Delay K  \\
\hline
$M_{ZZ} > 0.5$ & 1.0 & 11 & 6.2 & 5.2 & 1.1 & 1.5 & 2.5 & 2.1 & 1.5
\\
$M_{ZZ} > 1.0$ & 0.3 & 4.8 & 3.4 & 2.3 & 0.5 & 0.8 & 1.7 & 1.3 & 0.8
\\
$M_{ZZ} > 1.5$ & 0.1 & 0.6 & 0.2 & 0.5 & 0.1 & 0.3 & 0.9 & 0.6 & 0.3
\\[2mm]
\hline
$W^+ W^+$ & Bkgd. & SM & Scalar & $O(2N)$ & Vec 2.0 & Vec 2.5 & LET
CG &
   LET K & Delay K  \\
\hline
$M_{\ell\ell} > 0.25$ & 3.5 & 6.4 & 8.2 & 7.1 & 7.8 & 11 & 25 &
21 & 15  \\
$M_{\ell\ell} > 0.5$ & 1.9 & 3.8 & 5.0 & 4.5 & 4.5 & 7.2 & 20 &
16 & 11  \\
$M_{\ell\ell} > 1.0$ & 0.3 & 0.7 & 0.7 & 1.1 & 0.6 & 1.5 & 8.3 &
5.8 & 5.3  \\[2mm]
\hline\hline
\end{tabular}
\end{center}
\end{table*}

\onehead{DISCUSSION}

The results in Table 2 summarize the outcome of our
study.  As expected, the signal rates are largest in
the resonant channels.  Note, however, that the
rates are all rather low.  The events are clean, but
the low rates will make it difficult to isolate
high-mass resonances.  We must be ready for this
worst-case scenario and leave the door open
for a high-luminosity program at the SSC.

A second conclusion from Table 2 is that all channels are
necessary.  For example, isospin-zero resonances
give the best signal in the $W^+W^-$ and $ZZ$ channels, while
isospin-one resonances dominate the $W^+ Z$ channel.  The
nonresonant models tend to show up in the $W^+W^+$ final state,
so there is a complementarity between the different
channels.$^2$

A third conclusion is that we cannot cut corners.
Accurate background studies are crucial if we hope to
separate signal from background by simply counting rates.
We must also try to measure all decay modes
the $W$ and $Z$, including $Z \rightarrow \nu\bar\nu$ and
$W,Z \rightarrow jets$.  Finally, we must work to optimize the
cuts that are applied to each final state, with an eye to
increasing the signal/background ratio without affecting the
total rate.  All these considerations indicate that if the
Higgs is heavy, or if technicolor is correct, SSC studies
of electroweak symmetry breaking might
need a mature and long-term program before they give rise to
fruitful results.

\onehead{ACKNOWLEDGMENTS}

I would like to thank my collaborators V.~Barger, K.~Cheung,
J.~Gunion, T.~Han, G.~Ladinsky, R.~Rosenfeld and C.-P.~Yuan.
I would also like to thank S.~Dawson and G.~Valencia for many
conversations on the effective Lagrangian approach
to electroweak symmetry breaking.

%\vglue1pc % for pagebreaking


\begin{thebibliography}{99}

\bibitem{lqt} B.W.~Lee, C.~Quigg and H.~Thacker, \it Phys. Rev.
\rm D16, 1519 (1977).

\bibitem{chan} M.~Chanowitz, these proceedings.

\bibitem{otn} M.~Einhorn, \it Nucl. Phys. \rm B246, 75 (1984).

\bibitem{ccwz} S.~Weinberg, \it Phys. Rev. \rm 166, 1568 (1968);
S.~Coleman,
J.~Wess and B.~Zumino, \it Phys. Rev. \rm 177, 2239 (1969);
C.~Callan, S.~Coleman, J.~Wess and B.~Zumino, \it Phys. Rev. \rm 177,
2247 (1969); S.~Weinberg, \it Physica \rm 96A, 327 (1979).

\bibitem{BESS} R.~Casalbuoni, \it et al., Phys. Lett.
\rm B249, 130 (1990); B253, 275 (1991); J.~Bagger,
T.~Han and R.~Rosenfeld, Fermilab preprint CONF-90/253-T (1990).
See also M.~Bando, T.~Kugo and K.~Yamawaki, \it Phys. Rep. \rm
164, 217 (1988).

\bibitem{gasser} J.~Gasser and H.~Leutwyler, \it
Ann. Phys. \rm 158, 142 (1984); \it Nucl. Phys.
\rm B250, 465 (1985).

\bibitem{et}
M.~Chanowitz and M.K.~Gaillard, \it Nucl. Phys. \rm B261,
379 (1985).

\bibitem{bdv1} A.~Dobado, M.~Herrero and J.~Terron, \it
Z. Phys. \rm C50, 205 (1991); 465 (1991);
J.~Bagger, S.~Dawson and G.~Valencia, BNL
preprint 45782 (1990).

\bibitem{ewa} M.~Chanowitz and M.K.~Gaillard,
\it Phys. Lett.  \rm B142, 85 (1984);
G.~Kane, W.~Repko, and W.~Rolnick,
\it Phys. Lett. \rm B148, 367 (1984);
S.~Dawson, \it Nucl. Phys. \rm B249, 42 (1985).

\bibitem{bdv2} J.~Bagger, S.~Dawson and G.~Valencia, Fermilab
preprint PUB-92/75-T (1990), and references therein.

\bibitem{forward} R.~Cahn, S.~Ellis, R.~Kleiss and W.~Stirling,
\it Phys. Rev. \rm D35, 1626 (1987);
U.~Baur and E.~Glover, \it Nucl. Phys. \rm B347, 12 (1990);
V.~Barger, K.~Cheung, T.~Han and D.~Zeppenfeld, \it
Phys. Rev. \rm D44, 2701 (1991).

\bibitem{central} V.~Barger, K.~Cheung, T.~Han and R.~Phillips, \it
Phys. Rev. \rm D42, 3052 (1990);
D.~Dicus, J.~Gunion, L.~Orr and R.~Vega,  \it
Nucl. Phys. \rm B377, 31 (1992).

\end{thebibliography}
\end{document}